\documentclass[preprints,article,accept,moreauthors,pdftex]{mdpi}

\firstpage{1} 
\pubvolume{xx}
\issuenum{1}
\articlenumber{5}
\pubyear{2019}
\copyrightyear{2019}
\history{Received: date; Accepted: date; Published: date}

\pdfoutput=1

\usepackage{graphicx} 
\usepackage{amsmath}
\usepackage{mathtools}
\usepackage{amsfonts} 
\usepackage{anysize} 
\usepackage{subcaption} 

\graphicspath{{data/}}

		{\end{pmatrix}\end{medsize}}%

\newcommand\fig[1]{Fig.~\ref{#1}}

\newcommand\Wirr[0]{\langle W_{irr} \rangle}

\DeclarePairedDelimiter\abs{\lvert}{\rvert}%
\usepackage{color}   
\usepackage{hyperref}
\hypersetup{
	colorlinks=true, 
	linktoc=all,     
	linkcolor=blue,  
}

\newcommand{\mira}[1]{\vspace{5 mm}\par \noindent
	\marginpar{\textsc{acá}}
	\fcolorbox{red}{yellow}{\begin{minipage}[c]{0.95 \textwidth}
			\tt Cambie cosas en el párrafo de acá abajo \end{minipage}}
	\vspace{5 mm}\par
}

\usepackage{graphics} 
\usepackage{amsmath} 
\usepackage{amssymb}  

\Title{Driving interactions efficiently in a composite few-body system}

\Author{Alan Kahan$^{1,2}$, Thom\'as Fogarty$^{1,*}$, Jing Li$^{1}$ and Thomas Busch$^{1}$}

\AuthorNames{Firstname Lastname, Firstname Lastname and Firstname Lastname}

\address{%
$^{1}$ \quad Quantum Systems Unit, Okinawa Institute of Science and Technology Graduate University, Okinawa, 904-0495, Japan\\
$^{2}$ \quad Instituto de Física Enrique Gaviola, CONICET and Universidad Nacional de Córdoba, Ciudad Universitaria, X5016LAE, Córdoba, Argentina}

\corres{Correspondence: thomas.fogarty@oist.jp}

\abstract{We study how to efficiently control an interacting few-body system consisting of three harmonically trapped bosons. Specifically we investigate the process of modulating the interparticle interactions to drive an initially non-interacting state to a strongly interacting one, which is an eigenstate of a chosen Hamiltonian. We also show that for unbalanced subsystems, where one can individually control the different inter- and intra-species interactions, complex dynamics originates when the symmetry of the ground state is broken by phase separation. However, as driving the dynamics too quickly can result in unwanted excitations of the final state, we optimize the driven processes using shortcuts to adiabaticity, which are designed to reduce these excitations at the end of the interaction ramp ensuring that the target eigenstate is reached. 
}

\keyword{Shortcuts to adiabaticity; cold atoms; few-body systems}

\begin{document}


\section*{Introduction}

The ability to precisely control quantum systems is a pre-requisite for developing technologies in the areas of quantum computation, simulation and metrology. Even though the control over single particle states is highly developed by today \cite{lukin2003,lukin2013,lukin2018}, the requirements stemming from short decoherence time-scales are often hard to fulfil. Even more so, fast operations can have detrimental effects on quantum states as possible imperfections in the control pulses can be difficult to compensate. To mitigate these problems, and to ensure high fidelity on short timescales, a number of techniques have been developed, such as optimal control algorithms \cite{Werschnik_2007,optcont,nielsen2018} and shortcuts to adiabaticity (STA) \cite{torrontegui2013,delCampo2013,cui2016,sels2017,STA_review}. In this work we will focus on STAs, which are techniques designed to determine the driving parameters such that the system undergoes adiabatic evolution within a finite time, and which have been successfully employed in recent experiments with cold atoms \cite{schaff2010fast,schaff2011shortcut,schaff2011shortcuts,rohringer2015non,Deng2018,Diao_2018}. While for single particles or within a  mean-field approximation the description of the exact evolution of the quantum state is tractable, extending such a treatment to interacting many-body systems poses complications due to their complexity. One solution is to use approximate variational techniques, and, even though they are not exact, use these to design STA processes \cite{perez.garcia1996,Li2016,Li_2018}. In fact, it has recently been shown that such an approach can be used to control correlations in small systems \cite{lewis2019}, 
and can serve as a good benchmark for the control of larger interacting systems.

In this work we aim to move beyond mean-field and single particle physics to control the dynamics of larger systems with short-range contact interactions. To achieve this we take a first step by focusing on nontrivial few-body states, in particular controlling the interactions in a bosonic two-component system confined to a harmonic trap. Such systems can exhibit complex dynamics arising from the interplay of intra- and inter-species interactions, leading for example to composite fermionization or phase separation \cite{hall1998,mishra2007,garcia.march.2014,lee2016}. To discuss the basic effects we will focus on a paradigmatic realization of a two component system, namely two interacting ultracold atoms of species $A$ which interacts with a single ultracold atom of species $B$ in a one-dimensional setting. In such a system the interactions can be described by point-like potentials and in ultracold atoms experiments the interaction strength can be changed by employing Feshbach \cite{kohler2006,Chin2010} or confinement induced resonances \cite{olshanii1998}. In fact, individual tuning of the inter- and intra-species interactions allows one to explore driving the interactions of the $A$ atoms in the presence of an impurity atom, driving the interaction between an impurity and the interacting $A$ atoms, and also driving all interactions simultaneously. With this freedom it is possible to explore driving the system through the phase separation transition, which can alter the ordering of the particles in the trap and therefore greatly affect the dynamics of the system. We will show that efficient STAs can be designed for the individual interactions based on a variational ansatz, and that these STAs can outperform a non-optimized interaction ramp for most timescales of the driven process. 

In Sec.~\ref{Sec:model} we introduce the few-body model we consider and in Sec.~\ref{Sec:STA} we describe the variational method for designing STAs in this system. We begin in Sec.~\ref{Wirr.identical} with the analysis of driving interactions between three identical bosons, while in Sec.~\ref{Wirr_case2} and Sec.~\ref{strong.driving} we investigate driving one of the interaction terms while the other is held fixed. These latter sections describe the effect phase separation has on the driven system and its dynamics. Finally, we conclude our results.

\section{Model}
\label{Sec:model}
We consider a one-dimensional system of three interacting bosons of mass $m$ confined in a harmonic trap of frequency $\omega$, whose Hamiltonian is given by
\begin{equation}\label{eq:hamiltonian}
H= \sum_{j=1}^3\left[ -\frac{\hbar^2}{2m} \nabla_j^2  + \frac{1}{2} m \omega^2 x_j^2  \right]+ V_\text{int}(x_1,x_2,x_3)\;.
\end{equation}
At low temperatures we can assume that the scattering between the particles is mostly two-body and of s-wave form, allowing us to approximate the interaction part of the Hamiltonian with point-like pseudo-potentials as
\begin{equation}
V_\text{int}(x_1,x_2,x_3)  \approx  g^A(t) \delta(x_1 - x_2)+  g^{AB}(t) \left[ \delta(x_1-x_3) + \delta(x_2-x_3) \right]\;.
\end{equation}
Here the interaction strength between two particles of species A is given by $g^A$ and the interactions strength between an A particle and the B particle is given by $g^{AB}$. Both couplings are related to the respective 
3D scattering lengths, $a_{3D}$, via $g=\frac{4\hbar^{2}a_{3D}}{md_{\perp}^{2}}\frac{1}{1-C\frac{a_{3D}}{d_{\perp}}}$ with the constant $C$ given by $C\approx 1.4603$ \cite{olshanii1998}. Here $\omega_{\perp}$ is the trap frequency in the transverse directions of a quasi-one dimensional harmonic trap of width $d_{\perp}=\sqrt{\hbar/m\omega_{\perp}}$ and one can see that control over the transverse trap frequency or the scattering length allows one to tune the interaction strengths. 
In the following we will use harmonic oscillator units by rescaling all spatial coordinates with $a_0 \equiv \sqrt{\frac{\hbar}{m \omega}}$, all interaction strengths in units of $\sqrt{2} a_0 \hbar \omega$, all energies in units of $\hbar \omega$ and time in units of $\omega^{-1}$.

The Hamiltonian given in \eqref{eq:hamiltonian} can be separated by introducing Jacobi coordinates
\begin{align}
	X & = (x_1 - x_2 )/\sqrt{2}\;, \\
	Y & = (x_1 + x_2)/\sqrt{6} - \sqrt{2/3} x_3\;, \\
	Z & = (x_1 + x_2 + x_3)/\sqrt{3}\;, 
\end{align}
where $X$ and $Y$ are two relative coordinates describing the positions of the three particles with respect to each other, while $Z$ describes the systems center-of-mass. The latter separates and in these new coordinates the Hamiltonian can be written as  $H=H_\text{com}(Z) + H_\text{rel}(X,Y)$, with 
\begin{align}
H_\text{com} &= -\nabla_Z^2 + \frac{1}{2}Z^2\;, \\
H_\text{rel} &= -\frac{1}{2} \left( \nabla_X^2 + \nabla_Y^2\right) + \frac{1}{2}\left( X^2 + Y^2 \right)
+  g^A\delta(X) +  g^{AB} \left[
\delta \left(-\frac{1}{2}X + \frac{\sqrt 3}{2} Y\right) + \delta \left(-\frac{1}{2} X- \frac{\sqrt 3}{2}Y \right) \right]\;.
\end{align} 
It is immediately clear that the center-of-mass part just describes a particle moving in a harmonic trap, for which the solutions are given by 
\begin{equation}
	\psi^{HO}_n(Z)= \pi^{-1/4} (2^n n!) H_n(Z) \exp{(-Z^2/2)}\;,
\end{equation}
where $H_n(Z)$ are the Hermite polynomials, and with energies $E_n=(n+1/2)$.
The relative Hamiltonian can be interpreted as describing the effectively two-dimensional motion of a harmonically trapped particle in the presence of three narrow barriers arranged with a $\pi/6$ angle between them, as depicted in Fig.~\ref{fig:STA_ramps}(a). Since the two $A$ atoms are identical, the wave-function has to remain unchanged under a reflection across the $X$ coordinate, however there is no constraint when swapping particle $B$ with one of the $A$ particles. Exact solutions exist for limiting cases of the interactions \cite{Zinner_2014,Zinner2015}, while in general this system can be solved effectively using exact diagonalization techniques \cite{march2014,Miguel2013,garcia.march.2014}. Indeed, for $g^A\neq g^{AB}$ phase separation can be observed, whereby either the particles of species $A$ are pushed to the edges of the trap, while particle $B$ is confined in the trap center, or vice versa \citep{Cazalilla2003,Alon2006,mishra2007,Zollner2008,Miguel2013}. Therefore the arrangement of the atoms in the trap is non-trivial and will be strongly affected by any change in their inter- or intra-species interactions. It makes this an ideal system in which to study all regimes of composite few-body dynamics and to develop useful quantum control techniques.

In the following we will concentrate on increasing the interaction strength in initially non-interacting and uncorrelated systems 
on time scales $t_f$, which are approximately comparable to the trap period. As the center-of-mass contribution does not depend on the interaction it will be unchanged during the interaction ramping process, and therefore only the dynamics of the relative wave-function $i \hbar \frac{\partial}{\partial t}\psi(X,Y;t) = H_{rel}(t) \psi(X,Y;t)$ needs to be considered. 

For very slow ramps of the interaction ($t_f\ggg1$), the dynamics can be considered to be adiabatic and the energy of the state after the process is equal to the energy of the eigenstate at the final value of the interaction strength, $g_f$, i.e. $E^{AD}\equiv E(t_f)=E(g_f)$. For faster ramps ($t_f\sim\mathcal{O}(1)$) the process can be non-quasi-static resulting in excess energy in the system due to the out-of-equilibrium dynamics, with non-adiabatic energy $E^{NA}(t_f)\geq E^{AD}$. We can then define irreversible work from this non-adiabatic energy as
\begin{equation}
  \langle W_{irr} \rangle = E^{NA}(t_f)-E^{AD}\;,
\end{equation}
which for adiabatic processes will vanish, while being finite for non quasi-static processes \cite{lewis2019,garcia.march2016}. It is therefore a useful quantity to characterize the efficiency of the interaction ramp.


In the following we will describe two kinds of interaction ramps. The first will be a generic, non-optimised reference ramp which we will use as a benchmark 
and which we parameterize as
\begin{equation}
  g^\text{ref}(t)=\frac{g_f}{32}\left[ 30 \sin \left( \frac{\pi}{2} \frac{t}{t_f}\right) 
                                     -5  \sin \left( \frac{3\pi}{2} \frac{t}{t_f}\right)
                                      -3 \sin \left( \frac{5\pi}{2} \frac{t}{t_f}\right)                                              
                                              \right]\;.
\end{equation}
This function satisfies the boundary conditions $g^\text{ref}(0)=0$, $g^\text{ref}(t_f)=g_f$, and ${\dot{g}^\text{ref}}(0)={\dot{g}^\text{ref}}(t_f)={\ddot{g}^\text{ref}}(0)={\ddot{g}^\text{ref}}(t_f)=0$, ensuring that the interaction ramp begins and ends smoothly in an effort to reduce untypical excitations. However, this  assumption alone will not ensure that no irreversible dynamics are created during the ramping processes, and the pulse can therefore serve as a reference to an optimised interaction ramp derived from the STA approach. 
This comparison will be our main tool to quantify the success of the designed STA, which we will develop in the next section.




\section{Shortcut to adiabaticity}
\label{Sec:STA}

\begin{figure}[tb]
\centering
    \includegraphics[width=0.75\textwidth]{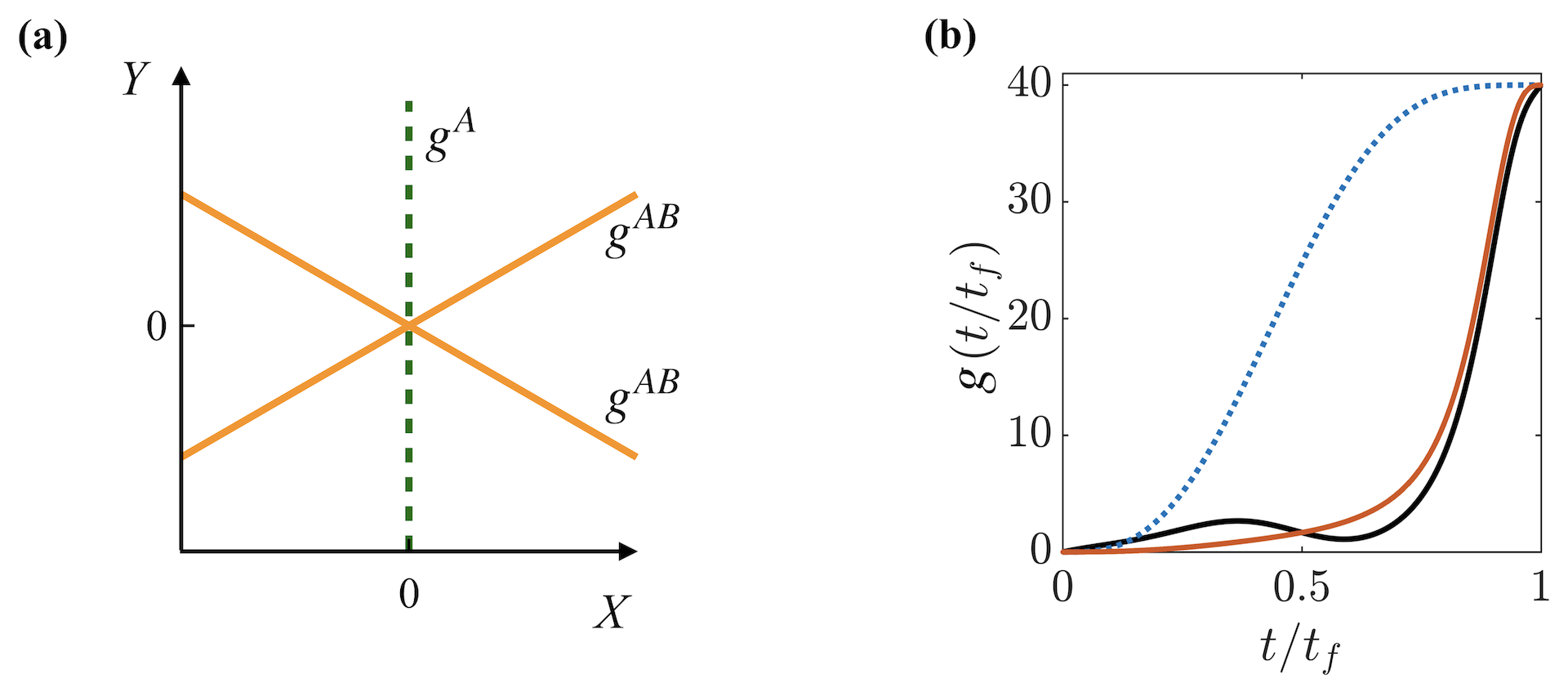}
  \caption{(a) Interaction potentials stemming from $g^{A}$ (dashed line) and $g^{AB}$ (solid lines) in the Jacobi coordinate plane.  (b) Examples of different interaction ramps when keeping $g^{A}(t)=g^{AB}(t)$, with the reference ramp (blue dotted line) and the STA ramp at $t_f=1.5$ (black solid line) and $t_f=10$ (red solid line).}
  \label{fig:STA_ramps}
\end{figure}

To find an STA, we use the method of inverse engineering which can design interaction ramps $g^\text{STA}(t)$ that fulfil the desired adiabatic evolution of the system $\phi(X,Y;t)$ for any ramp time $t_f$. The success of the STA then depends entirely on how well the solutions to the time-dependent Hamiltonian, $\phi(X,Y;t)$, are known, and if they possess scale invariance \cite{deffner2014}. If exact forms of the solution are not known, one has to use approximate techniques \cite{Li2016,andersen2016,zinner2017}, and here we employ a variational approach with an ansatz that describes the evolution of the relative wave-function by an interpolation between the initial and the final state 
\begin{align}\label{ansatz}
	\phi(X,Y;t)&= \varphi(X,Y;t)e^{i\left(b(t) X^2 + c(t) Y^2 \right) }\;,\\
	               &= N(t)\left[ (1-\eta(t)) \phi_i(X,Y) + \eta(t) \phi_f(X,Y) \right]e^{i\left(b(t) X^2 + c(t) Y^2 \right) }\;.
\end{align}
Here $N(t)$ is a time dependent normalization constant, while $b(t)$ and $c(t)$ are chirps that allow the wave-function to change its width. 
To ensure that the wave-function changes smoothly from the initial state $\phi_i(X,Y)\equiv \phi(X,Y;g_i)$
to the target state $\phi_f(X,Y)\equiv \phi(X,Y;g_f)$, we choose $\eta(t)$ as a $6^{th}$ order polynomial satisfying the boundary conditions
$\eta(0)=0$, $\eta(t_f)=1$, and $\dot{\eta}(0)= \ddot{\eta}(0) = \dot{\eta}(t_f) = \ddot{\eta}(t_f)=0$.

The next step in the process is to minimize the action of the effective Lagrangian \cite{perez.garcia1996} 
\begin{equation}\label{lagrangian}
\mathcal L  = \int_{-\infty}^\infty dX  \int_{-\infty}^\infty dY \left[ \frac{i}{2} \left( \frac{\partial \phi}{\partial t} \phi^*
- \frac{\partial \phi^*}{\partial t} \phi \right) 
-\frac{1}{2} \left| \frac{\partial \phi}{\partial X} \right|^2
 -\frac{1}{2} \left| \frac{\partial \phi}{\partial Y} \right|^2 
 -V(X,Y;t) \left| \phi \right|^2 \right]\;,
\end{equation}
where  $V=\frac{1}{2}\left( X^2 + Y^2 \right)+g^A(t)\delta(X) +  g^{AB}(t) \left[
\delta \left(-\frac{1}{2}X + \frac{\sqrt 3}{2} Y\right) + \delta \left(-\frac{1}{2} X- \frac{\sqrt 3}{2}Y \right) \right]$ contains the trapping and the  interaction potentials. This effective Lagrangian is minimized with respect to 
the variational parameters $\eta(t)$, $b(t)$ and $c(t)$ and the resulting Euler-Lagrange equations give the evolution equations that will 
determine $b(t)$ and $c(t)$ and establish the time dependence of the two interaction strengths, $g^{A}(t)$ and $g^{AB}(t)$, given by
\begin{equation}\label{EL3}
  g^A(t) I_A + g^{AB}(t) I_{AB}= - \left[ \frac{\partial \xi^2}{\partial \eta } \left( \dot b + 2 b^2 + \frac{1}{2}\right)
                                  + \frac{\partial \nu^2}{\partial \eta } \left( \dot c + 2 c^2 + \frac{1}{2}\right)
                                  + \frac{1}{2} \frac{\partial}{\partial \eta} \left( \beta + \gamma \right)
  \right]\;.
\end{equation}
Here $  \xi^2 = \int_\infty^\infty dX \int_\infty^\infty dY \; X^2 \abs{\varphi}^2$ and $\nu^2 = \int_\infty^\infty dX \int_\infty^\infty dY \; Y^2 \abs{\varphi}^2$ are the widths of the state in the $X$ and $Y$ directions respectively, and their dynamics is determined by
\begin{align}\label{EL1}
  \dot \xi &= 2 b \xi,\\
\label{EL2}
  \dot \nu &= 2 c \nu.
\end{align}
Similarly, the kinetic energies in these directions are $\beta = \int_\infty^\infty dX \int_\infty^\infty dY \left| \frac{\partial \varphi}{\partial X} \right| ^2$ and $\gamma = \int_\infty^\infty dX \int_\infty^\infty dY  \left| \frac{\partial \varphi}{\partial Y} \right| ^2$ and the interaction energies are $I_A = \int_\infty^\infty dX \int_\infty^\infty dY   \varphi \delta(X)$ and $I_{AB} = \int_\infty^\infty dX \int_\infty^\infty dY  \varphi \left[ \delta \left(-\frac{X}{2} + \frac{\sqrt{3}}{2}Y \right)+ \delta \left(-\frac{X}{2} - \frac{\sqrt{3}}{2}Y \right) \right]$. Using known solutions to $\phi_i$ and $\phi_f$ allows us to calculate these terms exactly (for example in the limit of infinite repulsive interactions accurate approximations are known \cite{Zinner_2014,Zinner2015}), otherwise these integrals can be calculated numerically. Typical examples of these STA interaction ramps for a system of three identical particles are shown in Fig.~\ref{fig:STA_ramps}(b) for two timescales $t_f=\lbrace 1.5,10 \rbrace$. When $t_f$ is large the STA is designed to quickly ramp the interaction at the end of the process, in contrast to the reference which begins to slowly increase the interaction already at the beginning of the ramp. This difference is due to the optimization of the STA through the Lagrangian, which is designed about the energy dependence on the interaction, and is similar to that seen in smaller systems \cite{lewis2019}. For small $t_f$ the STA possesses large modulations as driving the system faster requires large changes in the energy to follow the adiabatic path. 


\section{Three identical particles}\label{Wirr.identical}

\begin{figure}[htpb!]
\centering
    \includegraphics[width=\textwidth]{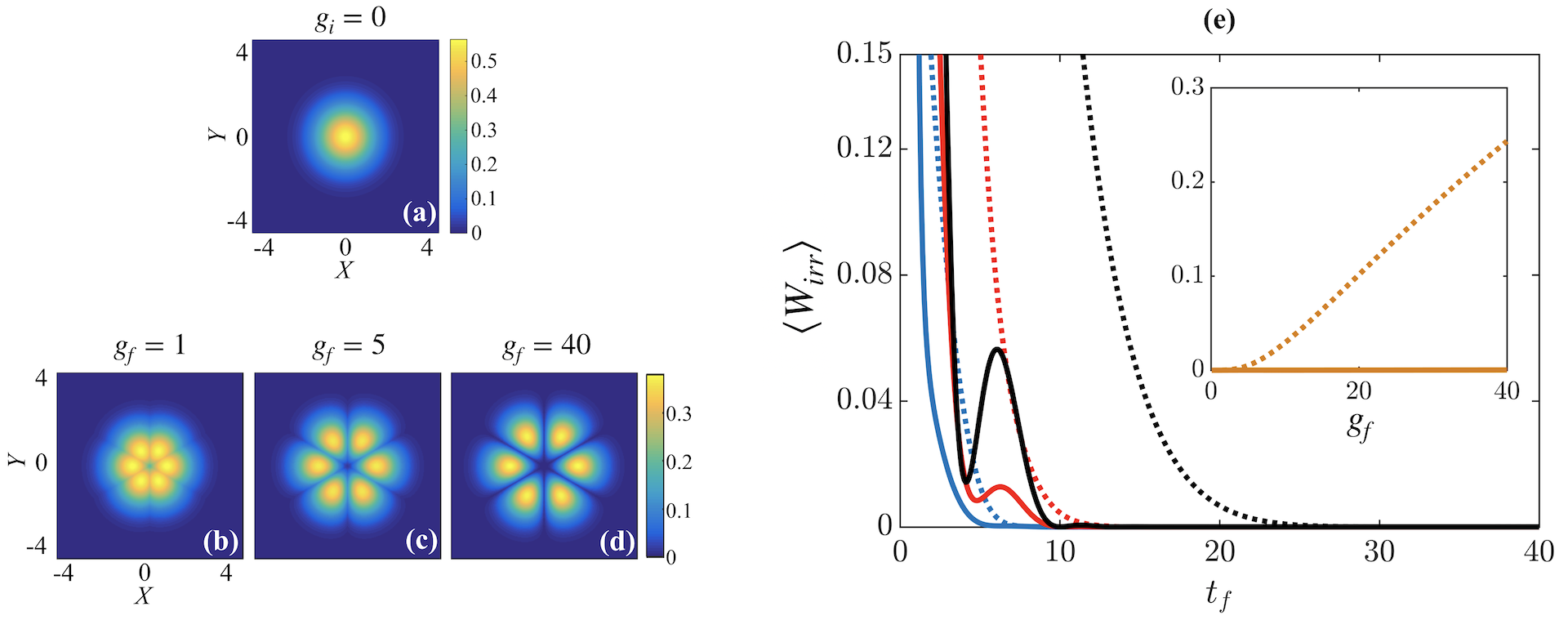}
  \caption{(a) Initial state in the relative $\lbrace X,Y \rbrace$ coordinate plane. Target states at (b) $g_f=1$, (c) $g_f=5$ and (d) $g_f=40$. (e) $\Wirr$ for three indistinguishable particles as a function of the ramp time $t_f$ for the STA (solid lines) and reference ramp (dotted lines). The final interactions are $g_f=1$ (blue lines), $g_f=5$ (red lines) and $g_f=40$ (black lines). Inset shows  $\Wirr$ versus final interaction strength $g_f$ at $t_f=10$.}
  \label{fig:case1_jacobi}
\end{figure}

Assuming that the initial state is the ground state of the noninteracting system, $g_i=g_i^A=g_i^{AB}=0$ (see Fig.~\ref{fig:case1_jacobi}(a)), we investigate in the following the ramping of strong interactions between three identical particles, i.e. $g(t)=g^A(t)=g^{AB}(t)$. In this case the relative part of the wave-function will always possess $C_{6\nu}$ symmetry \cite{Harshman2012,march2014,Harshman2016} (see Fig.~\ref{fig:case1_jacobi}(b-d)) with the interactions leading to cusps in the density at $60^{\circ}$ angles to each other, with the cusp asymptotically reaching zero density in the Tonks-Girardeau limit of strong repulsive interactions ($g_f\gtrsim 40$) \cite{kinoshita2004,minguzzi2005,haller2009}.

To quantify the success of the interaction ramps we compare the irreversible work $\Wirr$ after the respective interaction ramps, see \fig{fig:case1_jacobi}(e). For longer ramp times, $t_f > 25$, the system is driven slowly enough that it can be considered as evolving adiabatically, which results in a vanishing $\Wirr$, and therefore a high fidelity process for any reasonable non-optimized reference ramp. However for short ramp times, $t_f\lesssim 10$, ramping to stronger interactions creates more irreversible dynamics as the system is driven further from its equilibrium, resulting in large $\Wirr$ and therefore low-fidelity final states. It is on these timescales that we see the advantages of using the STA as it outperforms the non-optimised reference ramp, possessing lower amounts of $\Wirr$ for the different final interactions. 
The modulations visible in $\Wirr$ for the STA are due to excitations of the system to high energy states which possess the same symmetry as the ground state. While the contribution of these excitations is small for long ramp times, when driving the system quickly the STA is unsuccessful in damping them due to its approximate form through the ansatz in Eq.~\eqref{ansatz}. Indeed, divergence of $\Wirr$ occurs for timescales $t_f \lesssim  1$, when the STA ramp becomes negative at certain time intervals and destabilizes the system. This sets a limitation on the operation of our STA to times $t_f>1$. 

While for ramps to weak interactions the STA and the reference give comparable results, the difference between them increases when driving to larger interactions (see inset of \fig{fig:case1_jacobi}(e)). When using the STA the final state is essentially reached with $t_f=10$ for any final interaction, however applying the reference ramp drives the system further from the target eigenstate with growing interactions. This is not surprising as driving quickly to an infinitely repulsive state results in a diverging energy expectation value \cite{Kormos2014,Schmelcher2018}, however in this case the STA allows us to drive the system significantly faster than the reference as the excitations are successfully suppressed.


\begin{figure}[htpb!]
    \includegraphics[width=\textwidth]{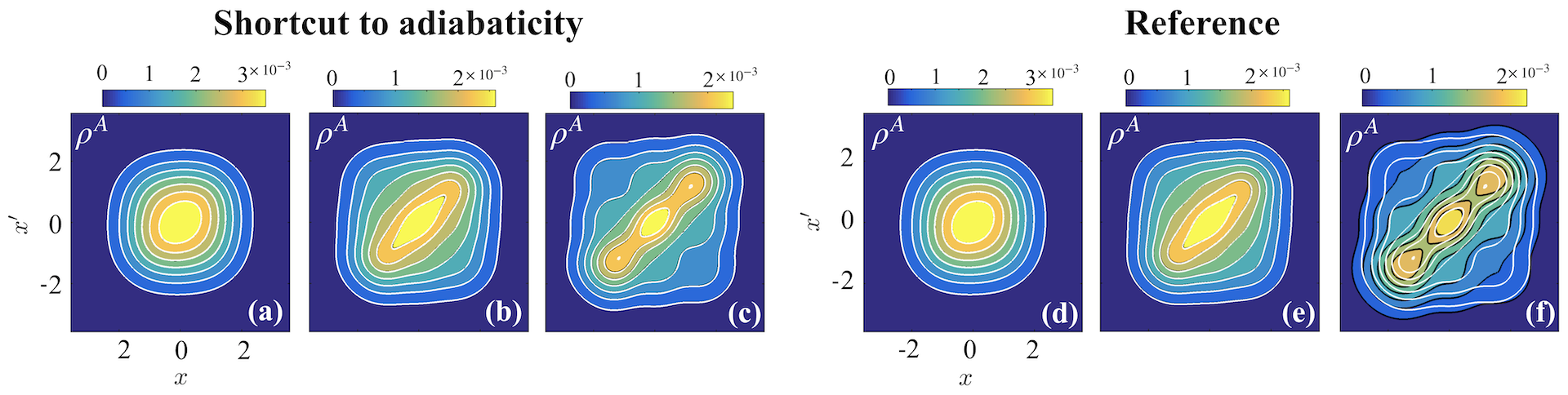}
  \caption{Target OBDMs (white contour lines) on top of final OBDMs for three identical particles at $t_f=10$.
  Panels (a-c) correspond to the STA, while panels (d-f) correspond to the
  reference pulse. Panels (a) and (d) are for $g_f=1$, (b) and (e) for $g_f=5$, and (c) and (f) for $g_f=40$.}
  \label{fig:case1_OBDM}
\end{figure}

Finally, we compare the structure of the three-body state through comparisons of the one-body density matrix (OBDM), whereby we examine the reduced state after tracing out two particles from the system
\begin{eqnarray}
\rho^A(x_1,x_1^{'})&=&\int_{-\infty}^{+\infty} \Psi(x_1,x_2,x_3) \Psi^*(x_1^{'},x_2,x_3) dx_2 dx_3\\
\rho^B(x_3,x_3^{'})&=&\int_{-\infty}^{+\infty} \Psi(x_1,x_2,x_3) \Psi^*(x_1,x_2,x_3^{'}) dx_1 dx_2\;.
\end{eqnarray}
Here $\rho^A(x_1,x_1^{'})$ is the OBDM of a particle of species A after tracing out the other A particle and the particle B, 
while $\rho^B(x_3,x_3^{'})$ is the OBDM of particle B after tracing out the two A particles. In the case when $g^A=g^{AB}$ both of these reduced states are equivalent, however this is not necessarily true when $g^A\neq g^{AB}$ as the components will be distinguishable and the rotational symmetry of the ground will be broken.

In Fig.~\ref{fig:case1_OBDM} we compare the OBDM after the STA and reference interaction ramps (at $t_f=10$) with that of the target OBDM. While ramps to weakly interacting states ((a,d) $g_f=1$ and (b,e) $g_f=5$) yield exactly the same OBDM after either of the ramps, the results are different when driving to strong interactions ((c,f) $g_f=40$). Indeed, the OBDM after the STA is equivalent to the target OBDM, however the OBDM after the reference is markedly different. Here it is instructive to compare the diagonal density ($\rho(x=x')$) of the reference OBDM to the target OBDM, which indicates that it has a broader width. In fact, each particle localizes spatially when the interactions are strongly repulsive, resulting in three density modulations about the center of the trap, however, the reference interaction ramp has imparted large kinetic energy to the system pushing the particles further from their equilibrium positions. In comparison the STA has precisely modulated the interaction to ensure that any non-adiabatic energy is removed from the system by the end of the ramp, resulting in a final state that is an eigenstate of the Hamiltonian. 


\section{Driving in the presence of weak fixed interactions}\label{Wirr_case2}

\begin{figure}[htpb!]
		\includegraphics[width=\textwidth]{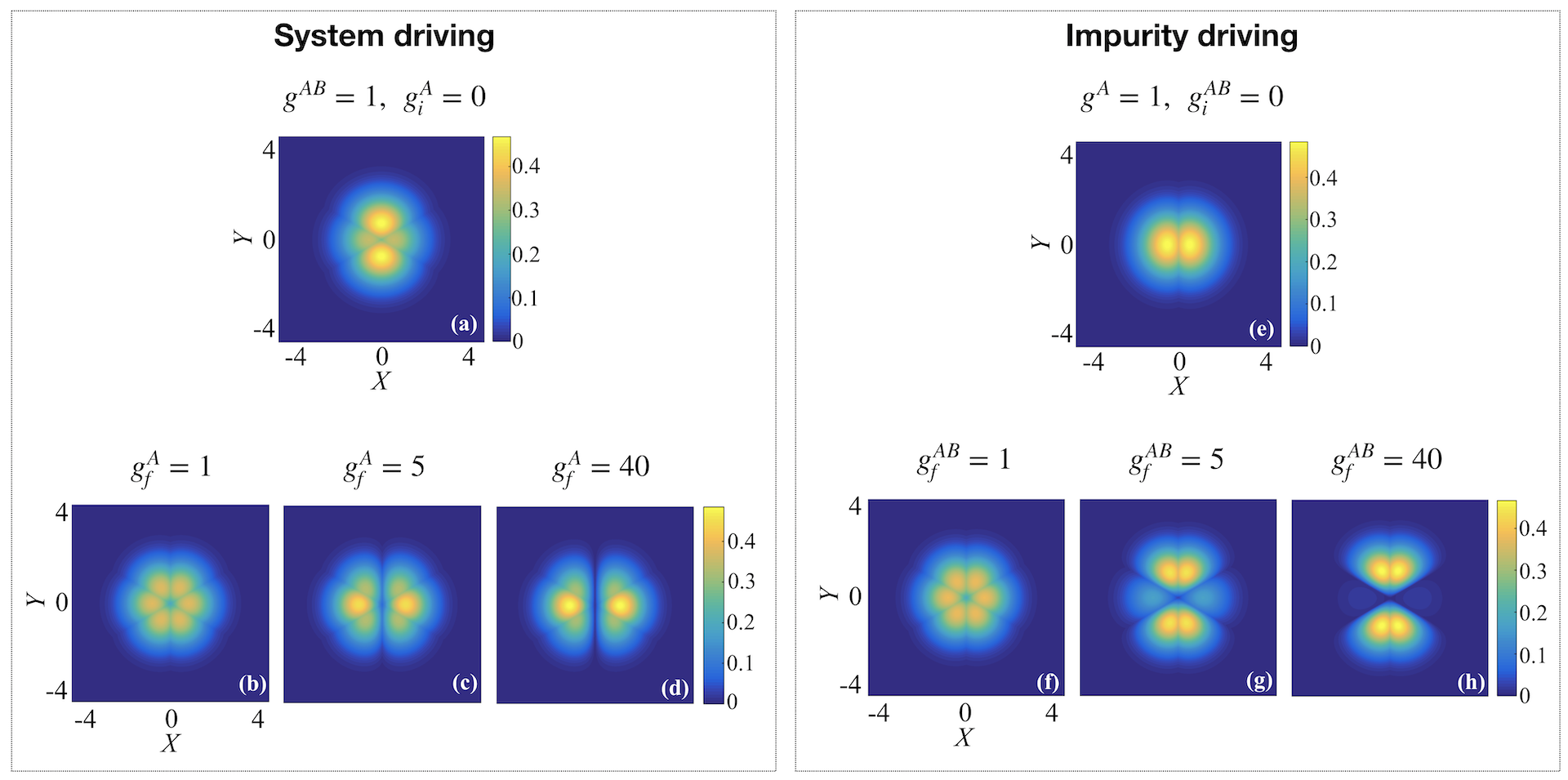}
		\caption{(a) Initial state with $g^{AB}=1$ and $g^A_i=0$, and (b-d) target states for $g^{A}_f=\lbrace 1,5,40 \rbrace$. This case we refer to as system driving. (b)  Initial state with $g^{A}=1$ and $g^{AB}_i=0$, and (b-d) target states for $g^{AB}_f=\lbrace 1,5,40 \rbrace$. This case we refer to as impurity driving.}
		\label{fig:weak_jacobi}
\end{figure}

Let us now consider that the inter- and intra-species interactions are different, such that we drive one interaction term while the other interaction term is held fixed at a low value. Then, due to the ability to tune both the inter- and intra-species interactions separately, one can consider two different setups. The first allows to tune the interactions between the A particles whilst in the presence of the impurity B atom, which we will refer to as system driving. At the same time the inter-species interaction between the impurity and the two A atoms is fixed at $g^{AB}=1$, such that the initial state at $g_i^{A}=0$ possesses small cusps along the directions $Y=\pm X/ \sqrt{3}$ (see Fig.~\ref{fig:weak_jacobi}(a)). Driving the interaction $g^{A}(t)>0$ between the A atoms will introduce a delta-function interaction potential which bisects the $X$-axis and force squeezing of the density from the $X$ to the $Y$ direction (see target states in panels (b) to (d)  in Fig.~\ref{fig:weak_jacobi}).

The second setup allows to switch on the interaction with an impurity atom in a weakly interacting two-body system, which we will refer to as impurity driving. For this we consider the intra-species interaction to be always fixed at $g^A=1$, while the interaction with the B atom is initially $g_i^{AB}=0$. The initial state therefore describes two weakly interacting $A$ bosons with the interaction bisecting the $X$-axis of the relative part of the wave-function (see Fig.~\ref{fig:weak_jacobi}(e)). Driving the interactions $g^{AB}(t)>0$ between the impurity and the A atoms will introduce delta-function interaction potentials along $Y=\pm X/ \sqrt{3}$ and force the density to be squeezed in the $X$ direction when the fixed interaction between the A particles is weaker than the impurity interactions (see target states in panels (g) and (h) in Fig.~\ref{fig:weak_jacobi}).

\begin{figure}[htpb!]
\centering
    \includegraphics[width=\textwidth]{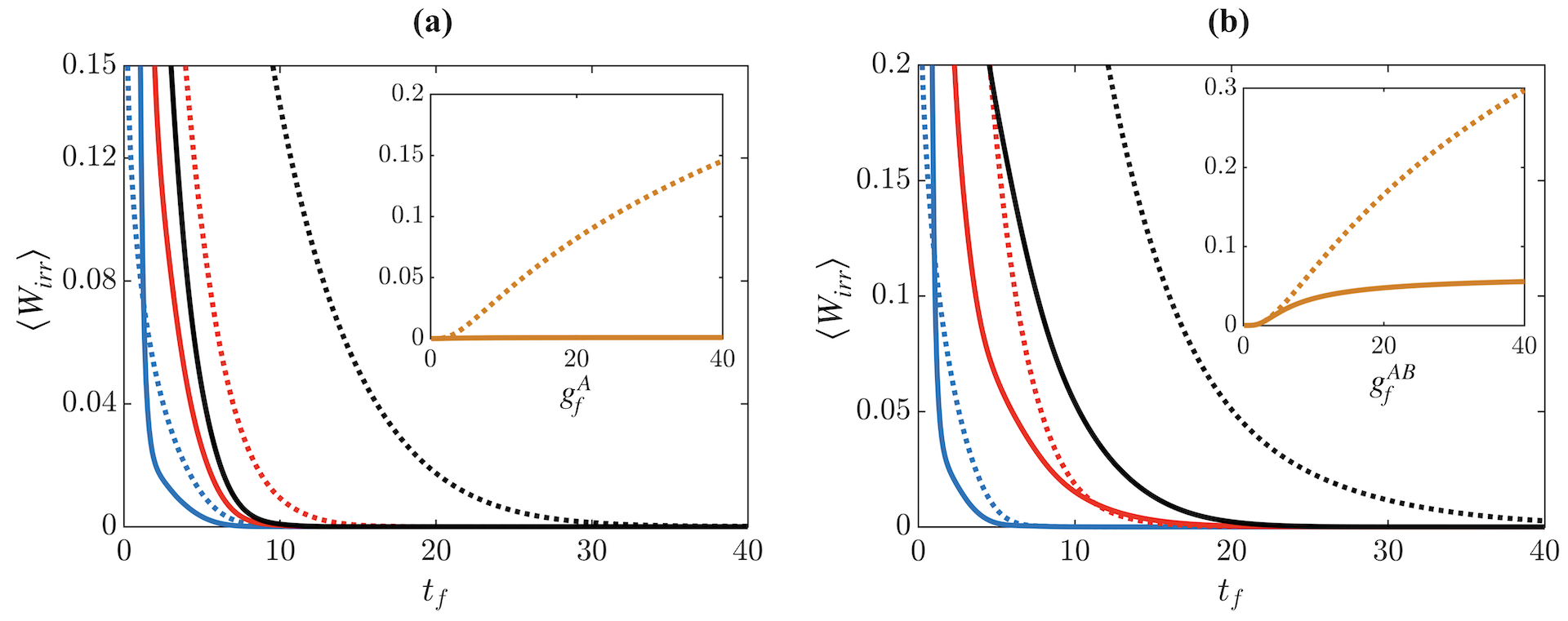}
  \caption{(a) $\Wirr$ after driving the system interactions in the presence of a weak fixed impurity interaction $g^{AB}=1$. System interactions are driven to $g^{A}_f=1$ (blue lines),  $g^{A}_f=5$ (red lines) and $g^{A}_f=40$ (black lines), with the solid lines showing the result of the STA and the dotted lines showing the result of the reference ramp. (b) Impurity driving in the presence of weak fixed system interactions $g_A=1$, with final impurity interactions $g^{AB}_f=1$ (blue lines),  $g^{AB}_f=5$ (red lines) and $g^{AB}_f=40$ (black lines). Insets show $\Wirr$ as a function of $g_f^A$ and $g_f^{AB}$ respectively at $t_f=10$.}
  \label{fig:case2_g1_Wirr}
\end{figure}

We assess the effect of the system driving dynamics through the irreversible work as shown in Fig.~\ref{fig:case2_g1_Wirr}(a). One can clearly see that for ramp times $t_f \geq 10$ the STA dynamics leads to the ground state irregardless of the strength of the final interaction $g^{A}_f$, while the reference ramp leads to increasing $\Wirr$ for stronger interactions on the same time scale (see inset in panel (a)). However, for times $t_f<10$ the STA becomes  less effective, similar  to the case of the simultaneous driving of $g^A$ and $g^{AB}$ discussed in Sec.~\ref{Wirr.identical} and the results found in \cite{lewis2019}.

For the case of driving the impurity we show  $\Wirr$  in Fig.~\ref{fig:case2_g1_Wirr}(b). The results here are subtly different to the previous case as the STA seemingly gives less of an advantage over the reference ramp when the interactions are weak, $g_f^{AB}\leq 5$. In fact, adiabaticity (when $\Wirr\rightarrow 0$) is reached for similar timescales for both the reference and the STA, and only when driving to large interactions, $g_f^{AB}=40$, the STA performs significantly better (see inset in panel (b)). 


\begin{figure}[h!]
\begin{center}
\includegraphics[width=\textwidth]{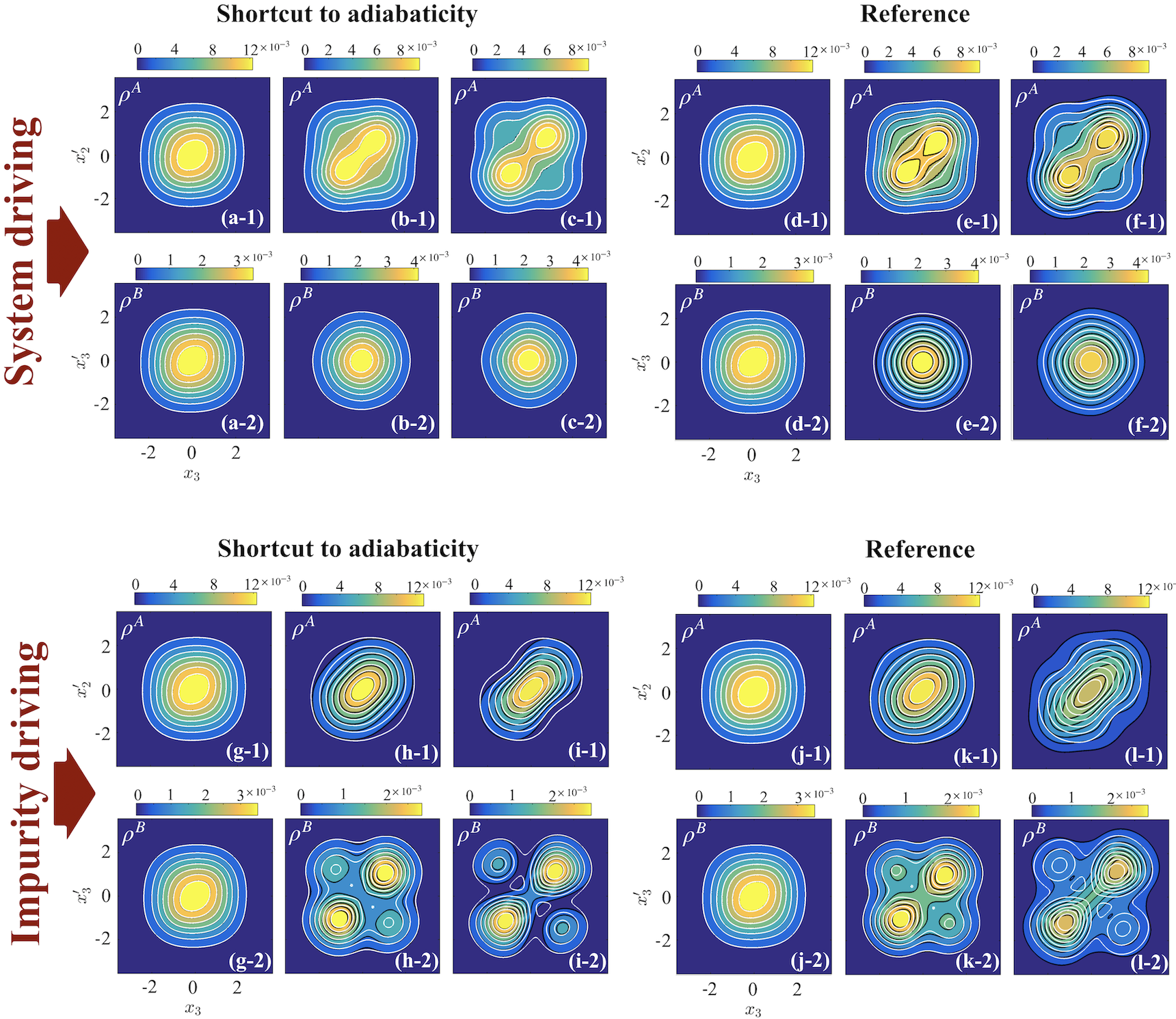}
\end{center}
  \caption{Panels (a-f): Target states (white contour lines) on top of final states (at $t_f=10$) for driving system interactions between A atoms in the presence of a fixed impurity interaction $g^{AB}=1$. Panels with index ($-1$) corresponds to $\rho^A(x_1,x_1^\prime) = \rho^A(x_2,x_2^\prime)$, while panels with index ($-2$) show $\rho^B(x_3,x_3^\prime)$. Panels (a-c) show the STA final states, while panels (d-f) show the final states for the reference pulse, where panels (a) and (d) are for $g^A_f=1$, (b) and (e) are for $g^A_f=5$, and (c) and (f) are for $g^A_f=40$. Panels (g-l): Target states (white contour lines) on top of final states (at $t_f=10$) for driving impurity interactions in the presence of a fixed interactions between the A atoms $g^{A}=1$. Final interactions are in the same order as above, panels (g) and (j) are for $g^{AB}_f=1$, (h) and (k) are for $g^{AB}_f=5$, and (i) and (l) are for $g^{AB}_f=40$.}
		\label{fig:OBDM_weak}
\end{figure}

The fact that driving the system and driving the impurity  produce qualitatively different results with regards to the creation of 
irreversible work can be explained by the need for a spatial 
re-organisation of the particles during the interaction ramp, which can be observed in the structure of the OBDMs for the system and the impurity 
(see Fig.~\ref{fig:OBDM_weak}). Firstly, driving the system interactions increases the repulsion between the two A particles, resulting in the appearance of two maxima in the diagonal density for the $A$ particles (see panels (b-1,c-1)), while the OBDM of the impurity B atom is relatively unchanged due to the weak inter-species interaction $g^{AB}=1$ (see panels (b-2,c-2)). Similar to Sec.~\ref{Wirr.identical} the STA successfully reaches the target eigenstate, while the reference forces the two A atoms further apart due to its inefficient driving (see (e-1,f-1)). In comparison, driving the impurity interaction forces the B atom to split and occupy the trap edges, being in a superposition of the left and right sides of the trap (see panels (h-2,i-2)). The strong inter-species interaction squeezes the density of the A atoms as they sit in the middle of the trap surrounded by the B particle (panels (h-1,i-1)). In this case the splitting of the impurity atom is a manifestation of phase separation in microscopic systems as the lighter species moves to the trap edges \citep{Alon2006,mishra2007,Zollner2008,Miguel2013}. Driving to a phase separated state is kown to lead to large irreversible dynamics as the particles oscillate between the miscible and immiscible regimes \cite{nicklas2015observation,bisset2015scaling,Mistakidis_2018}. Therefore the STA is less effective at driving the system quickly, however it can still outperform the non-optimised reference ramp (see panels (k) and (l)) bringing the final state closer to target eigenstate.


\section{Driving in the presence of strong fixed interactions.}\label{strong.driving}

\begin{figure}[htpb!]
\begin{center}
		\includegraphics[width=\linewidth]{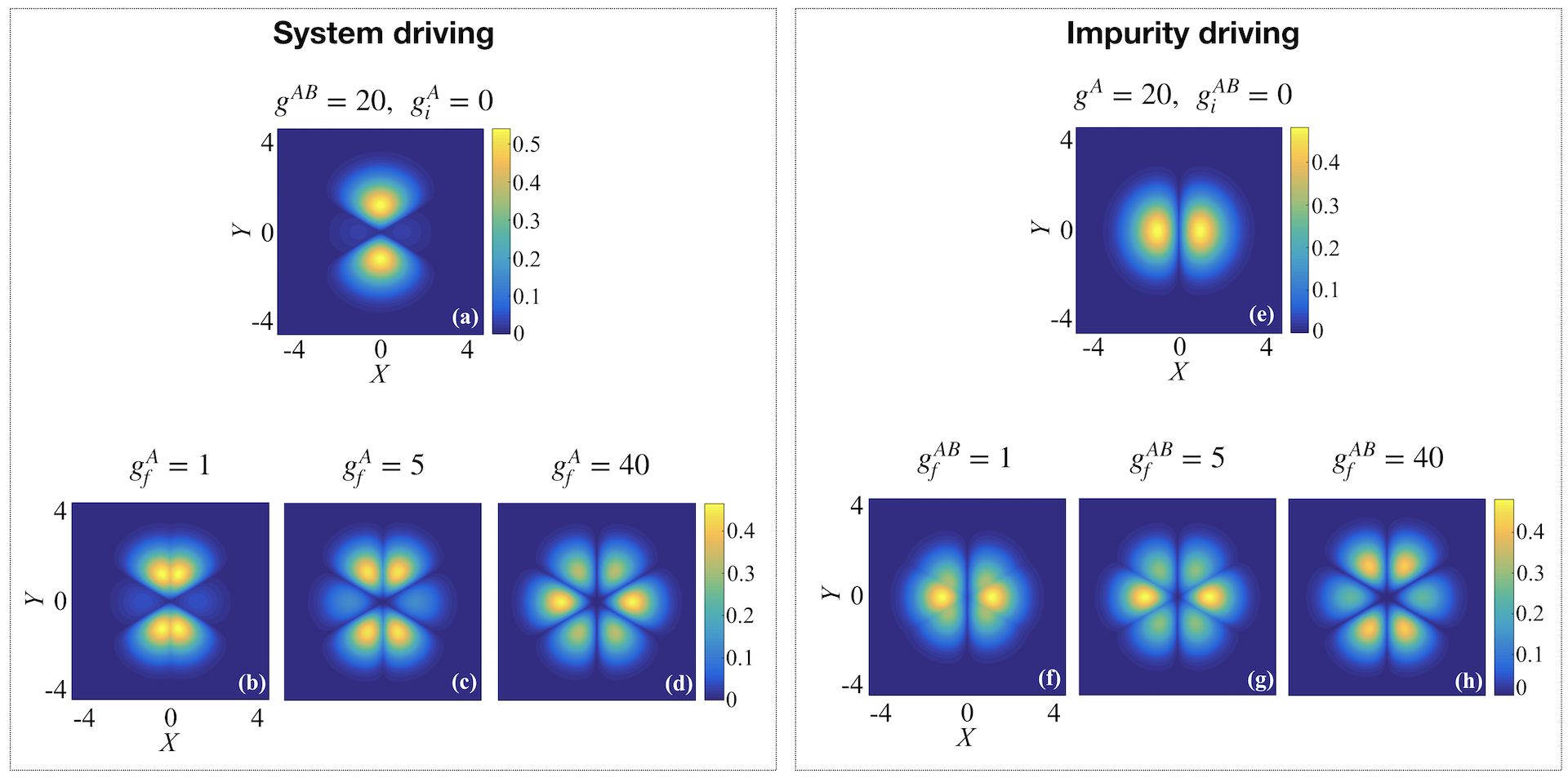}
		\end{center}
		\caption{Left panel: driving the interactions of the system when the impurity interaction is fixed at $g^{AB}=20$. (a) Initial state with $g^A=0$, (b) target state at $g^A_f=1$, (c) $g^A_f=5$ and (d) $g^A_f=40$. Right panel: driving the interaction with the impurity when the system interactions are fixed at $g^{A}=20$. (a) Initial state with $g^{AB}=0$, (b) target state at $g^{AB}_f=1$, (c) $g^{AB}_f=5$ and (d) $g^{AB}_f=40$.}
		\label{fig:JACOBI_strong}
\end{figure}

Let us next consider the driving of one interaction term while the other interaction term is held fixed at a large value, which we can again separate into two different setups. Driving the system interactions $g^A(t)$ when the impurity interaction is fixed at $g^{AB}=20$, and driving the impurity interactions $g^{AB}(t)$ when the system interactions are fixed at $g^{A}=20$. Similar to the previous section the presence of strong interactions in this system can result in the formation of different structural phases as the individual interaction terms are ramped. For instance, in the case of system driving a strong interaction with the impurity at $t=0$ forces the relative wavefunction density to align along the $X=0$ axis, while being suppressed along the $Y=0$ axis (see Fig.~\ref{fig:JACOBI_strong}(a)). Increasing the interactions among the A atoms then drives them apart along the $X$ direction (see panels (b) and (c)) and forces the density to redistribute to the previously unoccupied sectors between dips caused by the $g^{AB}$ interactions. At $g^A_f=40$ the orientation of the maximum density is rotated by $90^\circ$ from the initial state and lies along the $Y=0$ axis (see panel (d)). The inverse of this process is observed when driving the impurity interaction in the presence of strong $g^A$ (see panels (e-h)). This reorganisation of the particle density highlights the emergence of different phase separated regimes in the three-particle system. Specifically, for fixed impurity interactions $g^{AB}=20$ and for a final system interaction of $g^{A}_f=5$ the A particles are surrounded by the B particle (in the configuration BAB), while for $g^{A}_f=40$ the orientation is flipped with the B particle surrounded by the A particles (in the configuration ABA). Similarly, for fixed system interaction $g^A=20$ and $g^{AB}_f=5$ the B particle is surrounded by the A particles (ABA), and when $g^{AB}_f=40$ the A particles are surrounded by the B particle (BAB). Transitioning between the different phase separated regimes will become important when driving the dynamics.

\begin{figure}[tb]
\centering
    \includegraphics[width=\textwidth]{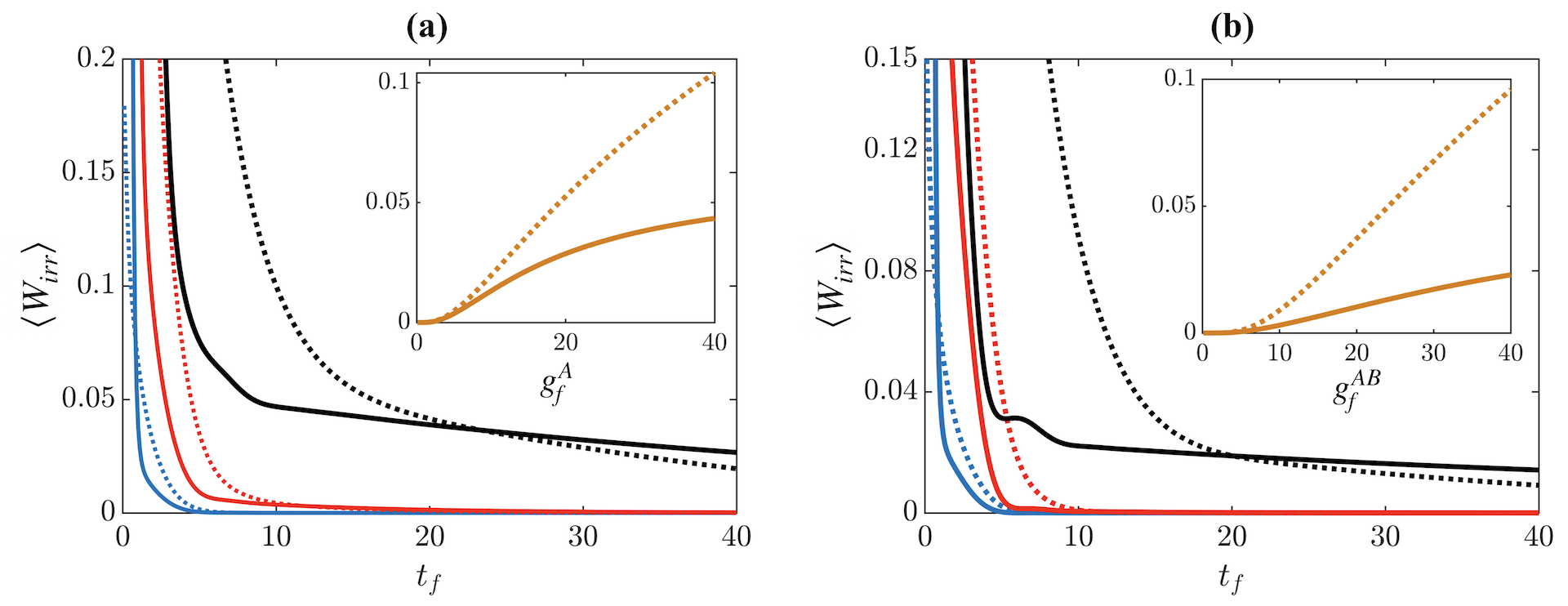}
  \caption{(a) $\Wirr$ after driving the system interactions in the presence of a strong fixed impurity interaction $g^{AB}=20$. System interactions are driven to $g^{A}_f=1$ (blue lines),  $g^{A}_f=5$ (red lines) and $g^{A}_f=40$ (black lines), with the solid lines showing the result of the STA and the dotted lines showing the result of the reference ramp. (b) Impurity driving in the presence of strong fixed system interactions $g^A=20$, with final impurity interactions $g^{AB}_f=1$ (blue lines),  $g^{AB}_f=5$ (red lines) and $g^{AB}_f=40$ (black lines). Insets show $\Wirr$ as a function of $g_f^A$ and $g_f^{AB}$ respectively at $t_f=10$.}
  \label{fig:Wirr_strong}
\end{figure}

\begin{figure}[tb]
    \includegraphics[width=0.9\linewidth]{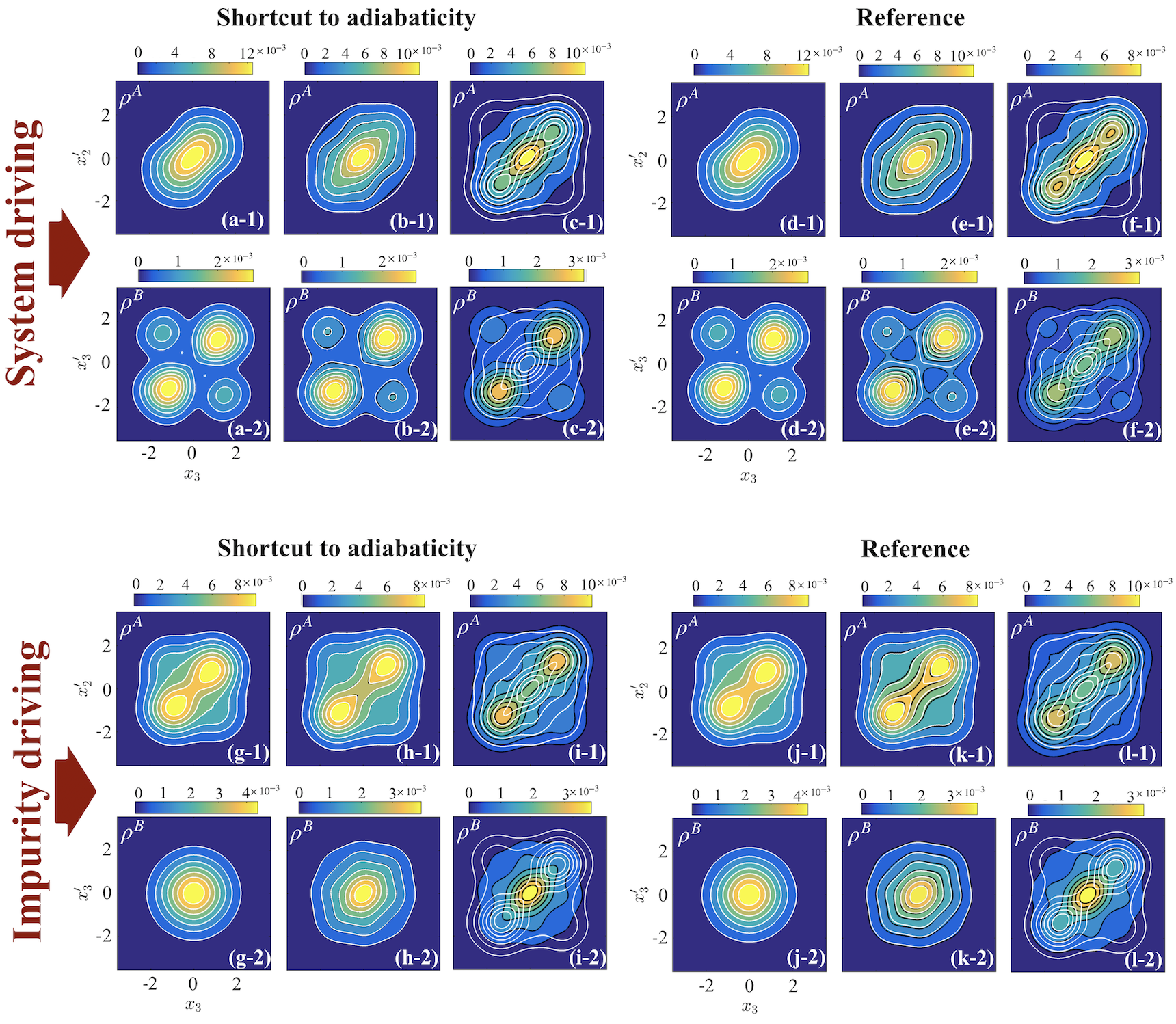}
  \caption{Panels (a-f): Target states (white contour lines) on top of final states (at $t_f=10$) for driving system interactions between A atoms in the presence of a fixed impurity interaction $g^{AB}=20$. Panels with index ($-1$) corresponds to $\rho^A(x_1,x_1^\prime) = \rho^A(x_2,x_2^\prime)$, while panels with index ($-2$) show $\rho^B(x_3,x_3^\prime)$. Panels (a-c) show the STA final states, while panels (d-f) show the final states for the reference pulse, where panels (a) and (d) are for $g^A_f=1$, (b) and (e) are for $g^A_f=5$, and (c) and (f) are for $g^A_f=40$. Panels (g-l): Target states (white contour lines) on top of final states (at $t_f=10$) for driving impurity interactions in the presence of a fixed interactions between the A atoms $g^{A}=20$. Final interactions are in the same order as above, panels (g) and (j) are for $g^{AB}_f=1$, (h) and (k) are for $g^{AB}_f=5$, and (i) and (l) are for $g^{AB}_f=40$.}
      \label{fig:OBDM_strong}
\end{figure}

In Fig.~\ref{fig:Wirr_strong} we show the irreversible work after driving the system interactions (panel (a)) and the impurity interactions (panel (b)). In both cases and for driving to weak final interactions, $g_f=1$ or 5, the STA outperforms the reference and the target state is reached on timescales $t_f\approx 10$. This is also confirmed by examining the OBDM at $t_f=10$ (see Fig.~\ref{fig:OBDM_strong}(a,b) and (g,h)), where the states after the STA closely match the target OBDM, while the states after the reference ramp (panels (d,e) and (j,k)) exhibit a slight mismatch at $g^f=5$. Indeed, since the strong fixed interactions dominate the structure of the three particle state, the positions of the particles relative to each other is not significantly altered which allows for efficient ramping of weak time-dependent interactions.

However, driving to stronger interactions, $g_f=40$, in both setups creates large amounts of irreversibility, irregardless of using the STA or the reference (see Fig.~\ref{fig:Wirr_strong}), with the STA producing less $\Wirr$ for ramp durations $t_f<20$. The insets of Fig.~\ref{fig:Wirr_strong} show $\Wirr$ at $t_f=10$ as a function of the final interaction strengths $g^{A}_f$ and $g^{AB}_f$, and while the rate of increase for the STA is less than that of the reference it is still significant. The reason for the failure of the different interaction ramps to reach the target state is the need for particles to spatially reorganise when the intra- and inter-species interactions become comparable (i.e.~when $g^{A}(t)\approx g^{AB}$ during system driving and $g^{AB}(t) \approx g^{A}$ during impurity driving). For example, when driving the system interactions, the relative wave-function of the initial state is highly localized along the $X$ direction due to the presence of the strong impurity interactions at $Y=\pm X/ \sqrt{3}$ (see Fig.~\ref{fig:JACOBI_strong}(a)), resulting in a phase separated state in the configuration $BAB$. Ramping to weak intra-species interactions ($g_f^{A}=\lbrace 1,5\rbrace$) preserves the phase separation as the impurity interaction still dominates the system (see Figs.~\ref{fig:OBDM_strong}(a,b) and (d,e)). Increasing the intra-species interaction further to $g^A=g^{AB}$ requires the ground state to regain the $C_{6\nu}$ symmetry discussed in Sec.~\ref{Wirr.identical}, however this is difficult to achieve dynamically due to the presence of the large fixed inter-species interactions, $g^{AB}=20$, which will suppress tunneling of the density across the delta-function barriers at $Y=\pm X/ \sqrt{3}$. In the three particle coordinates this corresponds to the need for the particles to tunnel through each other \cite{Pflanzer_2009,Pflanzer2010}, however they cannot reorganize on this short timescale due to the low tunneling rates. Therefore, for the final state of $g_f^{A}=40$ the particles will remain in the phase separated configuration of the initial state, $BAB$, instead of reaching the configuration of the target state, $ABA$ (see Fig.~\ref{fig:OBDM_strong} panels (c) and (f)). The same result is observed when driving the impurity interactions (see Fig.~\ref{fig:OBDM_strong} panels (i) and (l)) but with the opposite configuration of the particles.

\begin{figure}[tb]
    \includegraphics[width=\textwidth]{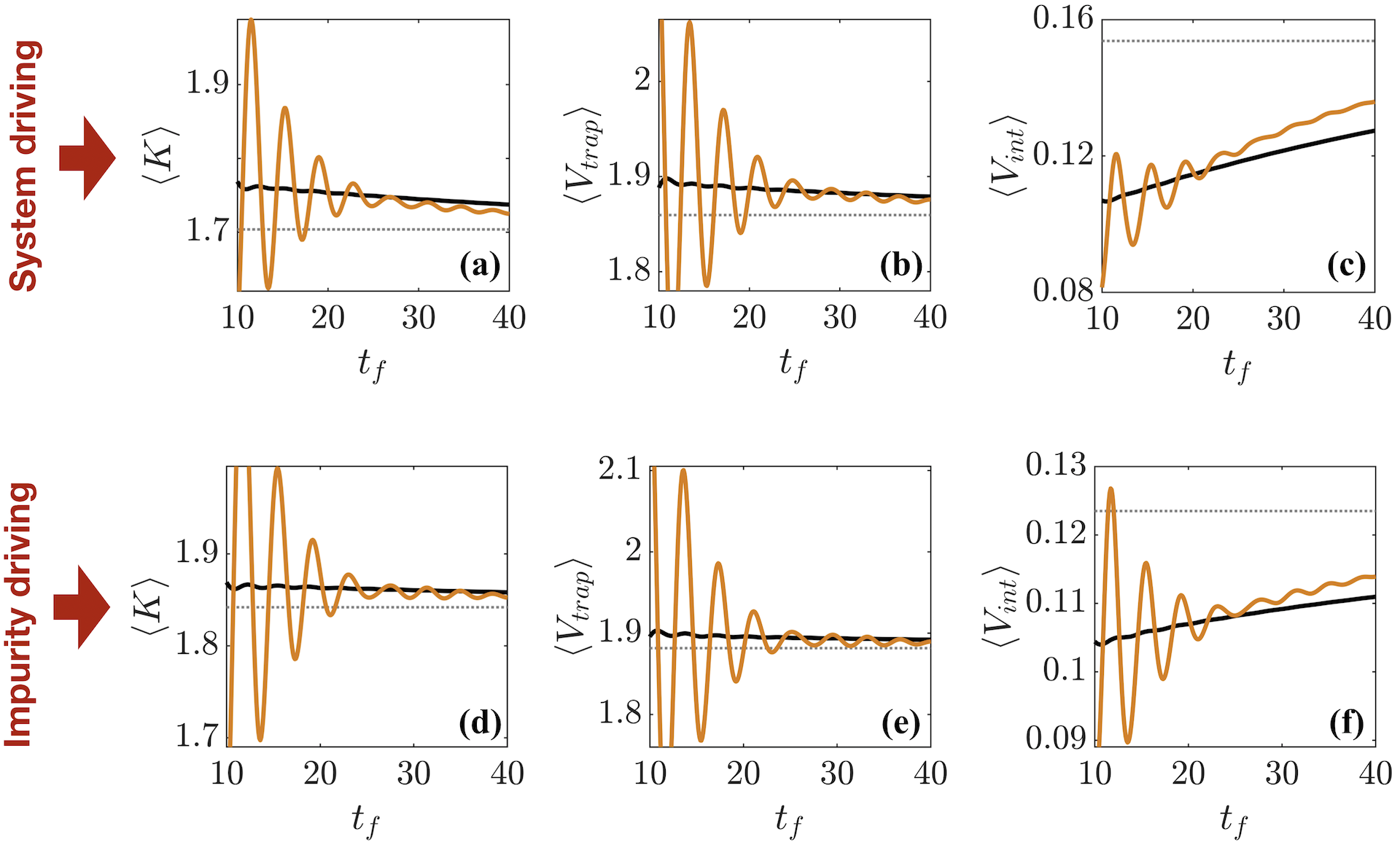}
  \caption{(a,d) Kinetic energy, (b,e) potential trap energy and (c,f) interaction energy as a function of $t_f$ after using the STA (black solid line) and reference (orange solid line), with the adiabatic energies shown as the thin dotted line. (a-c) Shows the result of driving the system to $g^A_f=40$, while (d-f) shows the result of driving the impurity to $g^{AB}_f=40$. Note the different scales on the subplots.}
  \label{fig:energies}
\end{figure}

Since these final states are not eigenstates, their energy exceeds that of the target ground state resulting in finite $\Wirr$ (see Fig.~\ref{fig:Wirr_strong}), even for long ramp times ($t_f\approx 40$) when we would assume the driving is slow enough to be close to adiabatic. Moreover, one can see that for $t_f>20$ the $\Wirr$ of the reference ramp is lower than that of the STA and that it decreases at a faster rate, suggesting that the higher energy states the reference ramp drives the system through prevent it from being fully trapped in the local energy minimum given by the wrong particle ordering. To understand this in more detail we show in Fig.~\ref{fig:energies} the average kinetic energy $\langle K \rangle$, trap potential energy $\langle V_{trap} \rangle$ and interaction energy $\langle V_{int} \rangle$ at the end of the different interaction ramps. One can immediately see that, for the STA (black lines), where the ramp is specifically designed to minimize excitations, only small oscillations of the energy components for any $t_f$ exist, while in comparison the reference ramp is not optimised to reduce excitations and therefore large oscillations are present. Furthermore, the kinetic and trap energies on average exceed that of the respective target states, whereas the interaction energy is less than that of the target state. This is consistent with the fact that both ramps are unable to achieve the correct ordering of the particles, therefore leaving them in an excited state that is more extended and therefore possesses slightly reduced interaction energies. It is also interesting to note that for long times ($t_f>20$) the reference ramp gives slightly more correct energies, which again is due it exciting higher lying states that help parts of the wave-function to not be trapped in the wrong state and encourages inter-particle tunneling.

\section{Conclusions \& Outlook}

We have presented a study that extends the use of shortcuts to adiabaticity to interacting few-particle systems. Using a variational technique we have shown that certain limits exist in which effective shortcuts for  these systems, which can possesses non-trivial groundstates, can be designed. Our approach works well for a number of cases, such as for three identical particles and in the presence of fixed weak interactions, and consistently outperforms a non-optimised reference ramp. The timescales for which high fidelity states can be reached are comparable to those achievable using optimal control techniques in related few-body systems \cite{nielsen2018}. We have also shown that the approach fails when the system is required to go through a phase-separation transition that requires the particles to tunnel through each other to achieve a new spatial ordering. 
As the STA is designed to reduce spurious excitations the particles become trapped in a quasi-stationary state with a lifetime much longer than the driving process. One way to enhance the tunneling rates of the particles during the interaction ramp would be to simultaneously modulate the trap potential \cite{muga2016}, apply periodic shaking of the trap \cite{Zenesini2010} or heat the system to allow for thermal hopping \cite{Wu2014}. These avenues will be explored in a future work. Also, while our superposition ansatz is effective for weak interactions (see Appendix \ref{App}), it can be improved to more accurately describe the phase separated regime by introducing intermediate states. Finally, while in this work we only consider a handful of particles it would be interesting to extend these techniques to larger systems where many-body effects such as the orthogonality catastrophe become significant \cite{Goold2011,Knap2012,Campbell2014,sels2017}.

\vspace{6pt} 



\authorcontributions{This project was conceptualized by TF, JL and TB. Software, validation, data curation and formal analysis by AK. Writing, review and editing by all authors.}

\funding{This work was supported by the Okinawa Institute of Science and Technology Graduate University. TF acknowledges funding support under JSPS KAKENHI-18K13507.}

\acknowledgments{We acknowledge support  from the Scientific Computing and Data Analysis section at the Okinawa Institute of Science and Technology Graduate University. AK acknowledges support and infinite patience from Cecilia Cormick.}

\conflictsofinterest{The authors declare no conflict of interest.}




\appendixtitles{yes} 
\appendix
\section{Accuracy of the ansatz}
\label{App}
\begin{figure}[htpb!]
	\includegraphics[width=\textwidth]{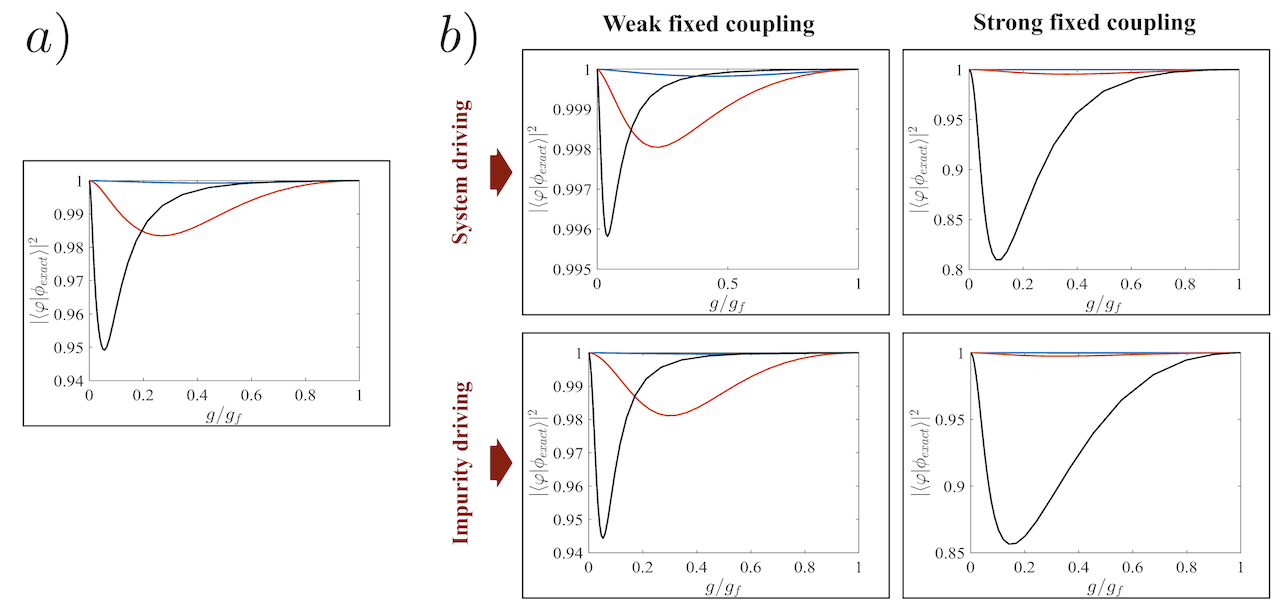}
	\caption{Fidelity between instantaneous ground state obtained with exact diagonalization of the Hamiltonian and
	the interpolatory ansatz as a function of $g/g_f$. Panel (a) is the fidelity for three identical particles while panel (b) shows the fidelity for system and impurity driving. The final interactions chosen are $g_f=1$ (blue lines), $g_f=5$ (red lines) and $g_f=40$ (black lines).}
	\label{overlap_si}
\end{figure}

To quantify the effectiveness of the interpolatory ansatz we use to design the STA ramps, we compare it with the exact eigenstates of the interacting Hamiltonian. 
In \fig{overlap_si} we show the fidelity as a function of the interaction strength, $|\langle \varphi (g)| \psi_{exact}(g)\rangle|^2$, where $\varphi(g)$ is the ansatz from \eqref{ansatz} and $\psi_{exact}(g)$ is the instantaneous eigenstate obtained through exact diagonalization.


In all cases when the final interaction strength is weak ($g_f=\lbrace 1,5 \rbrace $) the ansatz matches the exact eigenstate with almost unit fidelity. However, for strong fixed couplings and final interactions $g_f=40$ the fidelity drops significantly during the beginning of the interaction ramp as the ansatz fails to accurately account for the competition between the different interaction strengths. Similar deficiencies have been noted using other variational ansatzes for interacting few-body systems \cite{zinner2017}, and this can have a detrimental effect on the success of the STA in the presence of phase separation.

\reftitle{References}
\bibliography{biblio}{}

\end{document}